\documentclass[]{spie}  

\usepackage{amsmath,amsfonts,amssymb}
\usepackage{graphicx}
\usepackage[colorlinks=true, allcolors=blue]{hyperref}

\title{Thermal control of long delay lines in a high-resolution astrophotonic spectrograph}

\author[a]{Gregory P. Sercel}
\author[*, a, b]{Pradip R. Gatkine}
\author[a]{Nemanja Jovanovic}
\author[c]{Jeffrey B. Jewell}
\author[c]{Luis Pereira da Costa}
\author[c]{J. Kent Wallace}
\author[a]{Dimitri P. Mawet}
\affil[a]{Department of Astronomy, California Institute of Technology, 1200 E California Blvd, Pasadena, CA, 91125, USA}
\affil[b]{Department of Physics and Astronomy, University of California Los Angeles, 475 Portola Plaza, Los Angeles, CA, 90095, USA}
\affil[c]{Jet Propulsion Laboratory, California Institute of Technology, 4800 Oak Grove Drive, Pasadena, CA, 91109, USA}

\authorinfo{Further author information: (Send correspondence to G.P.S.)\\G.P.S.: E-mail: gsercel@caltech.edu\\ 
* NASA Hubble Fellow}

\begin{document} 
\maketitle

\begin{abstract}
High-resolution astronomical spectroscopy carried out with a photonic Fourier transform spectrograph (FTS) requires long asymmetrical optical delay lines that can be dynamically tuned. For example, to achieve a spectral resolution of R = 30,000, a delay line as long as 1.5 cm would be required. Such delays are inherently prone to phase errors caused by temperature fluctuations. This is due to the relatively large thermo-optic coefficient and long lengths of the waveguides, in this case composed of SiN, resulting in thermally dependent changes to the optical path length. To minimize phase error to the order of 0.05 radians, thermal stability of the order of 0.05° C is necessary. A thermal control system capable of stability such as this would require a fast thermal response and minimal overshoot/undershoot. With a PID temperature control loop driven by a Peltier cooler and thermistor, we minimized interference fringe phase error to +/- 0.025 radians and achieved temperature stability on the order of 0.05° C.  We present a practical system for precision temperature control of a foundry-fabricated and packaged FTS device on a SiN platform with delay lines ranging from 0.5 to 1.5 cm in length using inexpensive off-the-shelf components, including design details, control loop optimization, and considerations for thermal control of integrated photonics.  
\end{abstract}


\section{INTRODUCTION}
\label{sec:intro}  

The Fourier transform spectrograph (FTS) is a common and essential tool in spectroscopy, in which flux of interest is passed through a scanning interferometer, producing sinusoidal interference fringes \cite{zhang_research_2021}. As the delay path length in the interferometer is modulated, the measured optical power of the resulting fringes may be used to generate an interferogram. Taking the Fourier transform of the interferogram yields the optical spectrum of the input flux \cite{Gatkine_2019}. However, the resulting reconstructed spectrum is extremely susceptible to path length error within the interferometer; accurate and precise control of optical path length to a fraction of a wavelength is necessary. \cite{Paschotta2019fourier_transform_spectroscopy}

Fourier transform spectrographs have been demonstrated in integrated photonic platforms, in the conformation of Mach-Zehnder interferometers (MZIs), constructed on both silicon-on-insulator (SOI) and SiN-on-silica ($Si_3N_4/SiO_2$) platforms \cite{Nie:17, Souza_2018, xu_scalable_2024}.
Tunable optical path lengths in photonic FTS chips are possible due to the relatively high thermo-optic coefficients of these materials, however, this property results in a temperature-dependent optical path difference error. The optical path length of each arm of the interferometer, \textit{l}, is given by \(l = n\times d\), where \textit{n} is the refractive index of the waveguide and \textit{d} is the physical length of the waveguide.\cite{Kita2018} Assuming the reference arm and delayed arm of the interferometer possess different physical path lengths, the optical path difference between the two will vary with global chip temperature. Thus, precise thermal stabilization of photonic FTS chips is necessary to minimize phase error resulting from fluctuating optical path differences.\cite{Herrero-Bermello:17}

\begin{figure} [ht]
   \begin{center}
   \begin{tabular}{c} 
   \includegraphics[height=7cm]{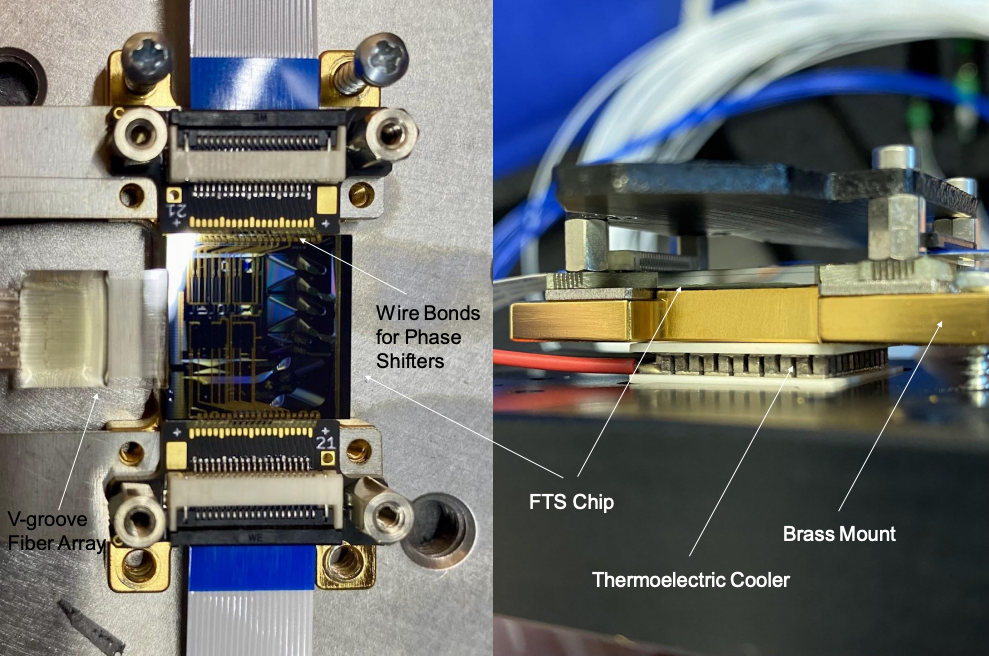}\end{tabular}
   \end{center}
   \caption[example] 
   { \label{Photos of Chips} 
Top view (left) and side view (right) of fully packaged photonic FTS chip mounted on thermoelectric cooling element.}
   \end{figure}

We present a thermal control system to minimize thermally induced phase errors in a photonic FTS chip fabricated with SiN waveguides, shown in Fig. \ref{Photos of Chips}, designed for astronomical spectroscopy.

\section{EXPERIMENTAL SETUP}

\subsection{Spectrograph Chip}
\label{sec:title}
\begin{figure} [h]
   \begin{center}
   \begin{tabular}{c} 
   \includegraphics[height=9cm]{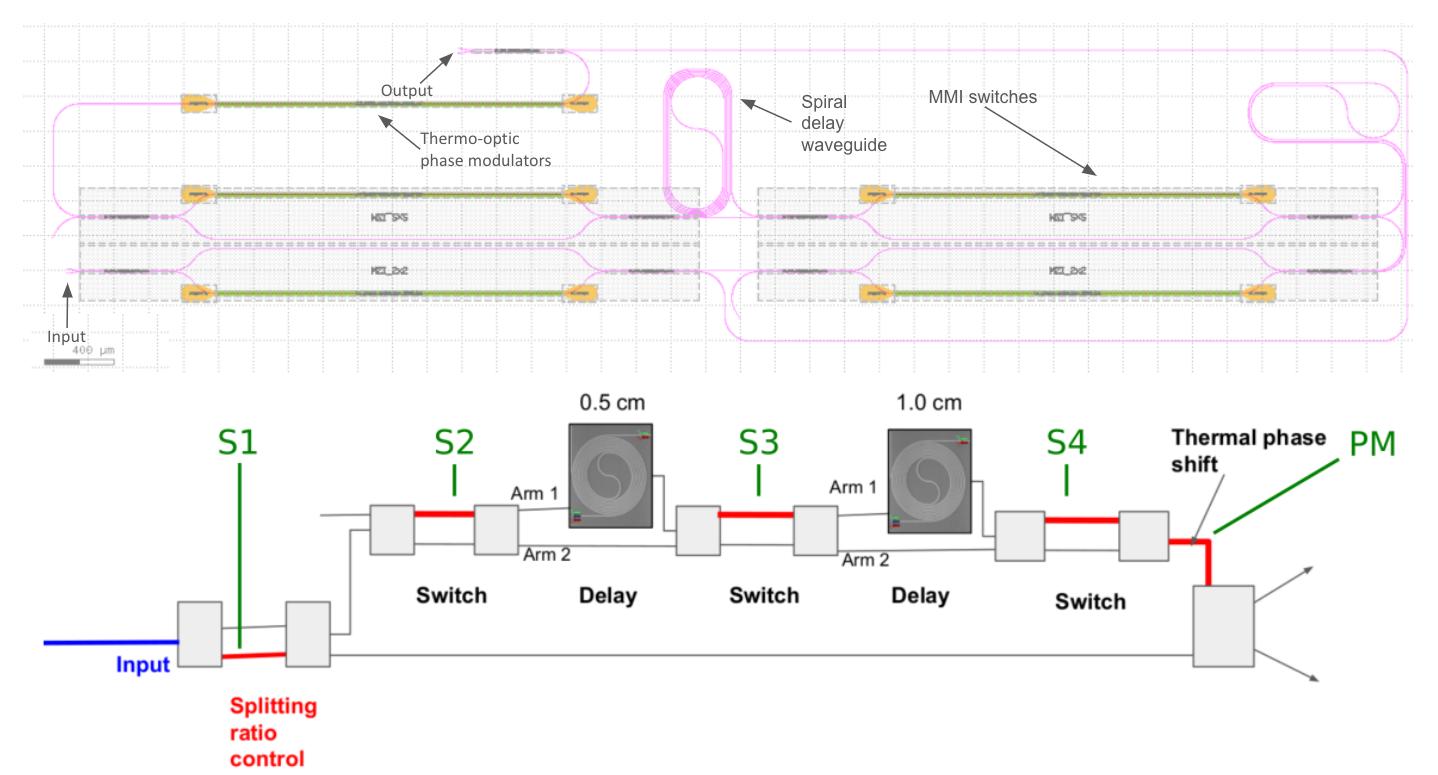}\end{tabular}
   \end{center}
   \caption[example] 
   { \label{fig:MZI Schematic and Mask Layout} 
\textbf{Top:} Main MZI mask layout depicting TOPMs (yellow), regions heated by TOPMs (green), and waveguides (pink). \textbf{Bottom:} Functional schematic depicting Archemedian spiral delay lines interspersed by MZI switches.}
   \end{figure}

The Fourier transform spectrograph chip shown in Fig. \ref{Photos of Chips} was designed by JPL, Caltech, and Bright Photonics, fabricated by LioniX International, and mounted and packaged with a V-groove fiber-optic pigtail by Phix Photonic Assembly. The FTS chip contains several large MZI structures, of which one is concerned in this paper. This large MZI, shown in Fig. \ref{fig:MZI Schematic and Mask Layout}, is composed of cascaded Archemedian spiral delay lines separated by smaller 2x2 MZIs labeled as switches that are constructed from multi-mode interferometers (MMIs) and thermo-optic phase modulators (TOPMs). To differentiate between the smaller MZI stages and the larger overarching MZI structure, the smaller MZIs will be called MZI switches, while the large cascaded structure will be called the main MZI.   

\begin{figure} [h]
   \begin{center}
   \begin{tabular}{c} 
   \includegraphics[height=4cm]{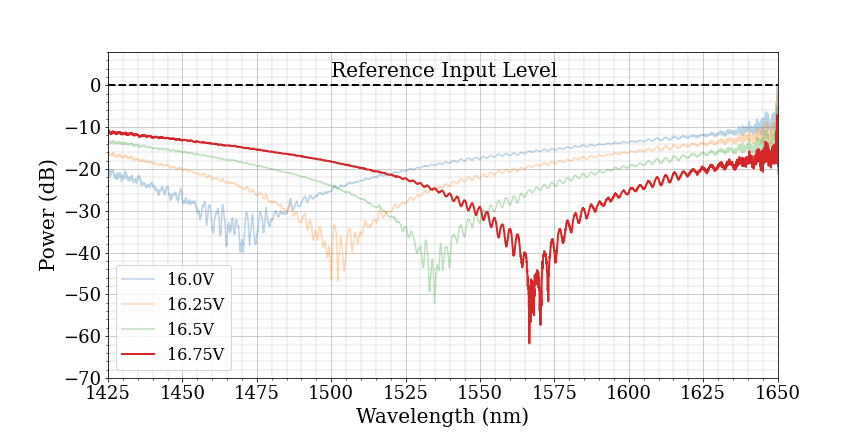}
\hspace{-0.75cm} 
\includegraphics[height=4cm]{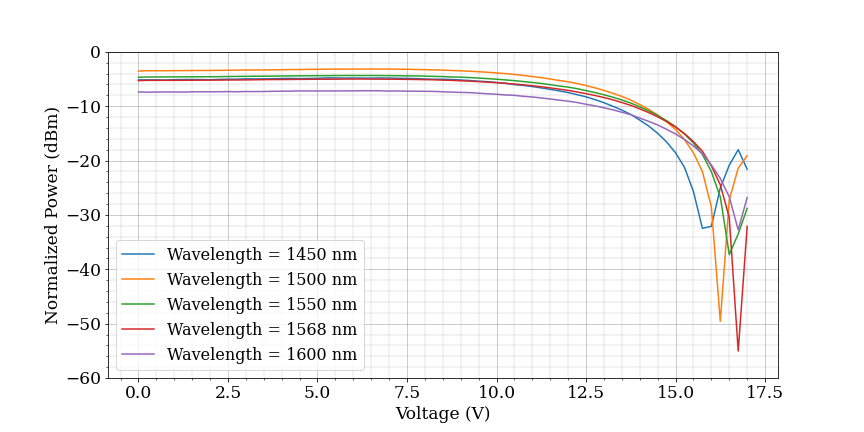}\end{tabular}
   \end{center}
   \caption[example] 
   { \label{fig:Plot Showing Broadband Suppression by MZIs of up to 50 dB} 
\textbf{Left:} Throughput of an MZI output as a function of wavelength, demonstrating broadband switchability and suppression of up to 50 dB around 1566 nm. \textbf{Right:} Throughput of the same MZI output as a function of voltage, showing suppression can be achieved over broad bands with voltages in the range of 16-17 V.}
   \end{figure}

Each MZI switch is used to control the transmission of optical power through each of their two outputs, which is accomplished via TOPMs.  Each TOPM is a thin chrome trace that runs along one waveguide within each MZI switch, to which a voltage bias is applied by a computer-controlled power supply. The resulting high current density heats the trace, thermally inducing a local change in the refractive index of the waveguide. Thus, through electrical control, flux may be directed through one output while suppressing flux at the other output, and vice versa. Figure~\ref{fig:Plot Showing Broadband Suppression by MZIs of up to 50 dB} demonstrates the suppression of flux by an MZI switch at several wavelengths. Large drops in throughput are seen at voltage biases of 16-17 V for all wavelengths, indicating large bandwidths can be controlled simultaneously.

The operation of an MZI is fairly simple: input flux is split down two paths, one of which contains a path delay while the other is a straight path, considered to be the reference path. Light traveling along the delayed path has a phase shift relative to the path without a delay, which can be modulated. The two paths are recombined and interfere constructively and destructively. By modulating the relative phase difference between the delayed path and reference path, an interference fringe can be reconstructed. The main MZI depicted in the functional schematic in Fig. \ref{fig:MZI Schematic and Mask Layout} follows this principle, where input flux enters a controllable MZI switch that splits light down both the delayed path and reference path, denoted as \textit{S1}. Then, additional MZI switches along the delayed path, denoted as \textit{S2}, \textit{S3}, and \textit{S4}, each are set to configure the optical path to either a straight waveguide which bypasses a delay, or an Archemedian spiral waveguide delay. By electrically setting switch states of \textit{S2}, \textit{S3}, and \textit{S4}, the length of the delayed path can be varied from 0 to 1.5 cm, in 0.5 cm steps. Finally, the delayed path is passed along a thermal phase shifter, denoted as \textit{PM}, that enables additional phase modulation, and then recombined with the reference path.

A stand-alone MZI switch is present on the chip for calibration purposes; it is used to determine the corresponding switching voltage for each optical path length configuration. The same experimentally determined switching voltages from this test structure are used to control the MZI switches within the main MZI.

\subsection{Thermal Control}
A thermoelectric cooling device (TEC) is used to control chip temperature. The FTS chip is fixed on a brass mounting plate which is thermally coupled to the cold side of a 30-watt TEC via thermal paste, as can be seen in Fig. \ref{Photos of Chips}. The TEC's hot side is thermally coupled to a steel cooling plate, from which heat is dissipated with an aluminum heat sink and cooling fan, as well as copper thermal straps fixed to an optical bench. The TEC is powered by a 5-volt supply with a 1-amp current limit through an N-MOS driver. The output power of the N-MOS is modulated with an Arduino Uno microcontroller via pulse-width modulation (PWM). 

\begin{figure} [ht]
   \begin{center}
   \begin{tabular}{c} 
   \includegraphics[height=7cm]{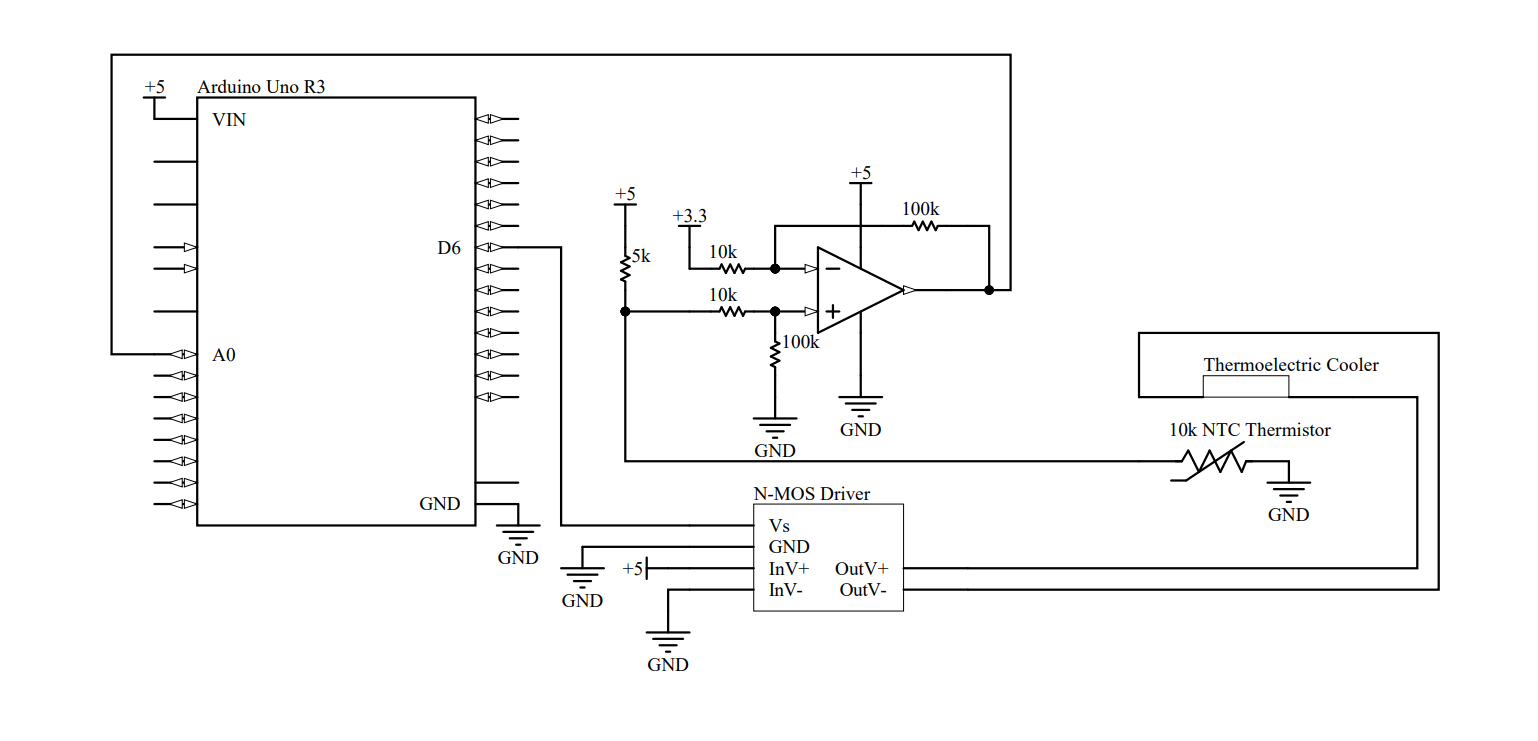}\end{tabular}
   \end{center}
   \caption[example] 
   { \label{fig:Electrical Layout} 
Electrical layout of the thermal control system.}
   \end{figure} 

A thermistor is fixed in a bore in the brass mounting plate underneath the FTS chip and is used for temperature feedback to drive a PID control loop. Analog voltage from the thermistor is amplified by a factor of 10 using a rail-to-rail op-amp, boosting the temperature sensitivity of the thermistor. This maintains temperature precision otherwise lost in quantization by the 10-bit analog-to-digital converter of the microcontroller. The electrical schematic of the entire thermal control apparatus is shown in Fig. \ref{fig:Electrical Layout}. 

PID control was implemented on the microcontroller using a built-in PID library, and the control loop has been manually tuned to achieve a fast settling time and minimal overshoot/undershoot using a PID tuning guide from \textit{Thorlabs}.\cite{pidtuning} An additional oversampling algorithm averages 10-sample bins in every loop cycle to mitigate the effects of electrical noise. The control loop maintains the chip at 18° Celsius, slightly below the ambient temperature of the laboratory. 

\subsection{Characterization}
\begin{figure} [ht]
   \begin{center}
   \begin{tabular}{c} 
   \includegraphics[height=3cm]{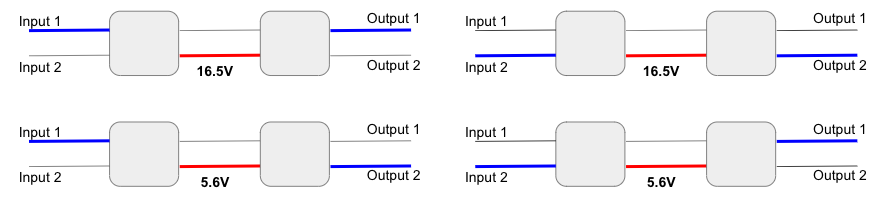}\end{tabular}
   \end{center}
   \caption[example] 
   { \label{fig:MZI Path Selection} 
Each MZI switch can be configured to transmit flux from the same side as the input waveguide at a 16.5 V bias, or the opposing side at a 5.6 V bias. }
   \end{figure}

\textbf{Identifying the switching voltages:} Using the aforementioned standalone MZI switch test structure, voltage biases corresponding to each switch configuration were experimentally determined by modulating the applied voltage from 0 V to 18 V in 0.1 V steps, while illuminating the device with an L-band monochromatic source and measuring optical power at both outputs with an optical power meter. The voltage biases associated with each switch state for the test structure were translated into truth tables for the main MZI, in which voltages applied to each of its switch stages will engage or disengage optical delay lines to set desired path lengths (see Fig. \ref{fig:MZI Path Selection}).

\begin{figure} [ht]
   \begin{center}
   \begin{tabular}{c} 
   \includegraphics[height=5cm]{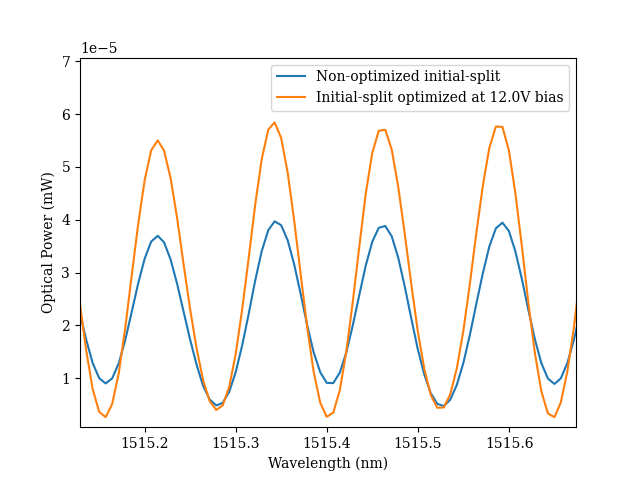}
\hspace{-0.75cm}
\includegraphics[height=4.5cm]{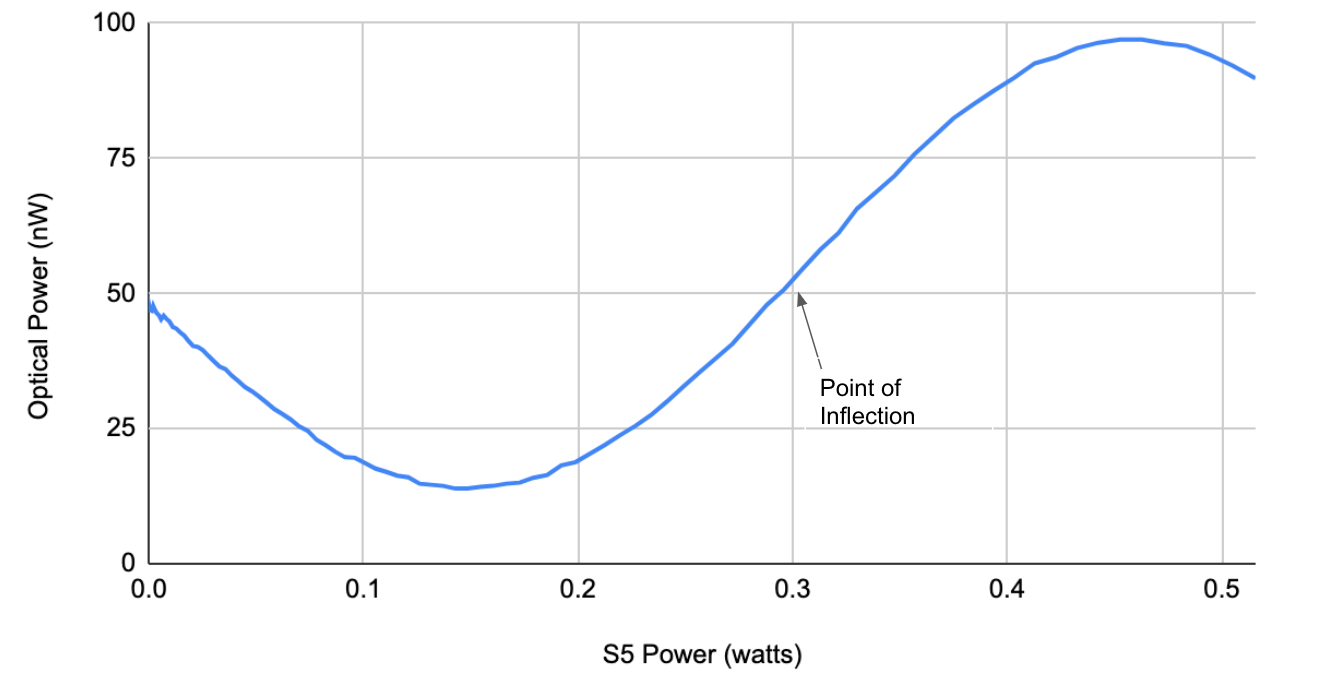}\end{tabular}
   \end{center}
   \caption[example] 
   { \label{fig:Demonstration of Fringe Amplitude and Phase Control} 
\textbf{Left:} Non-optimized and optimized sinusoidal interference fringes from MZI configured with a 1.5~cm delay path, demonstrating amplitude modulation by the initial splitting ratio MZI switch, \textit{S1}. \textbf{Right:} Optical power as a function of the voltage applied to the thermal phase shifter, \textit{PM}, at 1550 nm. }
   \end{figure}

\noindent \textbf{Balancing the power in reference and delayed arms:} Once each MZI path length configuration is set, the initial split-ratio-controlling MZI switch (\textit{S1} in Fig. \ref{fig:MZI Schematic and Mask Layout}) was incrementally modulated from 0 V to 18V in 0.25V steps while a 1550 nm super-luminescent diode (SLD) source was used to illuminate the device. The MZI output was passed to an optical spectrum analyzer (OSA) that was used to analyze the resulting interference fringes for each \textit{S1} voltage sweep. The experimentally determined \textit{S1} voltage that yields maximized interference fringe amplitude is considered to provide the optimal splitting ratio of flux between the delayed path and reference path (see Fig. \ref{fig:Demonstration of Fringe Amplitude and Phase Control}-left). This process was repeated for each delay path length configuration.

\noindent \textbf{Measuring the long-term phase error:} 
The MZI, once set to a given path length with an optimized splitting ratio, was fed with an L-band monochromatic source of constant flux, which was achieved with a variable optical attenuator. The wavelength-phase relationship of interference fringes, which may be approximated as sinusoids, is controlled by modulation of the \textit{PM} thermal phase shifter (see Fig. \ref{fig:Demonstration of Fringe Amplitude and Phase Control}-right). Assuming an amplitude \textit{A} for the sinusoidal fringe at given wavelength $\lambda$, we get, 
\begin{equation}
    \mathrm{Optical~power}~P  = A \times sin(\phi)
\end{equation}

\begin{equation}
    \delta P = A \times \delta\phi~cos(\phi)
\end{equation}

At the point of inflection of the sinusoidal fringe (i.e. where $sin(\phi)$ = 0) at a given $\lambda$, cos($\phi$) $\sim$ 1. Thus, the phase error ($\delta\phi$) can be inferred as:  
\begin{equation}
    \delta \phi = \delta P/A
\end{equation}


To experimentally determine A at a given $\lambda$, we modulated the MZI's thermal phase shifter to sweep over an interference fringe and measured the optical power at the fringe peak, trough, and point of inflection. After measuring the amplitude A, we set the MZI's thermal phase shifter to the point of inflection and leave it there for $\sim$ 2 hours while keeping the thermal control loop on. During this time, the optical power is continuously measured at a sampling rate of 15 Hz and used to compute phase error (using Eq. \ref{eq:fov}) as a metric of fringe stability. The phase error is calculated as
\begin{equation}
\label{eq:fov}
\delta\phi = \sigma/A\, ,
\end{equation}
where \(\sigma\) is the measured standard deviation in the optical power at the point of inflection of the fringe throughout the measurement (i.e. over 2 hours), and \textit{A} is the fringe amplitude measured at the beginning of the experiment.

\section{RESULTS AND DISCUSSION}

\subsection{Calibration}
The voltage biases found to direct all optical power from one input to each of the two outputs of the single-stage 2x2 MZI switch test structure, identical to those in each MZI variant, were found to be 5.6~V and 16.5~V, respectively. Figure \ref{fig:MZI Path Selection} depicts the optical path of a hypothetical MZI switch set to each corresponding voltage state, for a given input waveguide.

The \textit{S1} initial split-ratio MZI switch swept from 0 V to 18~V in 0.25~V steps for each path length configuration was found to produce maximized interference fringes at a 12~V bias in all cases. 

\subsection{Thermal Control}
After manually tuning PID coefficients to $K_{p}$, $K_{i}$, and $K_{d}$ values of 700, 25, and 1, respectively, the thermal control system demonstrated rapid settling and precise thermal stability at 18° C. 

\begin{figure} [ht]
   \begin{center}
   \begin{tabular}{c} 
   \includegraphics[height=4cm]{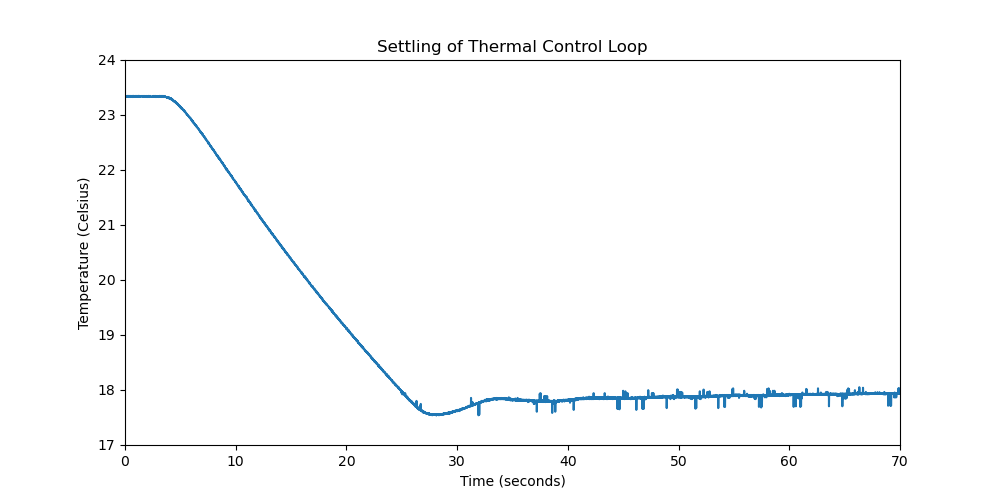}
\includegraphics[height=4cm]{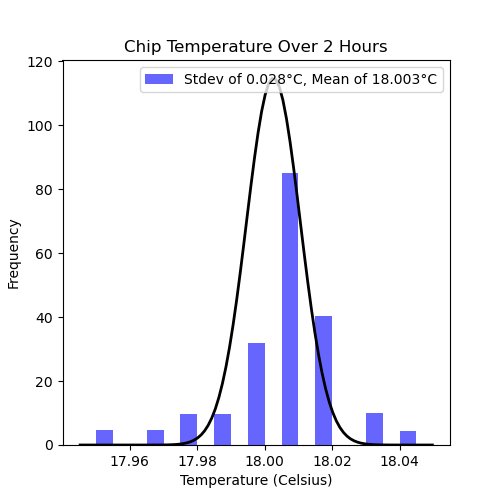}\end{tabular}
   \end{center}
   \caption[example] 
   { \label{fig:Thermal Stability and Settling Results} 
Settling curve of thermal control loop following activation (left), histogram of chip temperature over 2-hour monitoring period (right).}
   \end{figure} 

As the settling curve and histogram in Fig. \ref{fig:Thermal Stability and Settling Results} show, the device stabilizes to its set point from room temperature in under 60 seconds with minimal overshoot/undershoot once activated. Additionally, once stabilized, a standard deviation of +/- 0.028° C was achieved over a monitoring period of 2 hours. 

Once the thermal control system has been activated and stabilizes to the set temperature, the delay path length of the MZI is configured. This causes additional chip temperature swings which have been observed to settle in as little as 30 seconds. 

\subsection{Stability}
\begin{figure} [ht]
   \begin{center}
   \begin{tabular}{c} 
   \includegraphics[height=4cm]{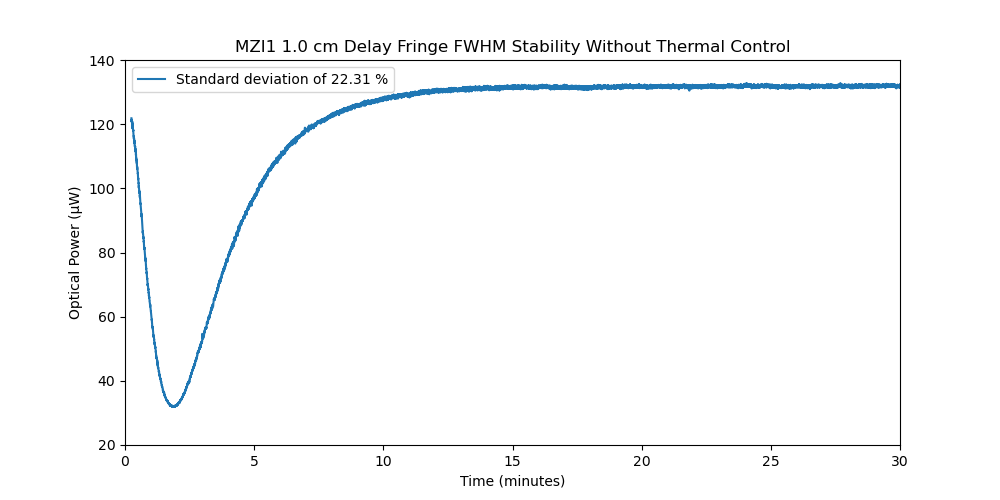}
\hspace{-0.75cm} 
\includegraphics[height=4cm]{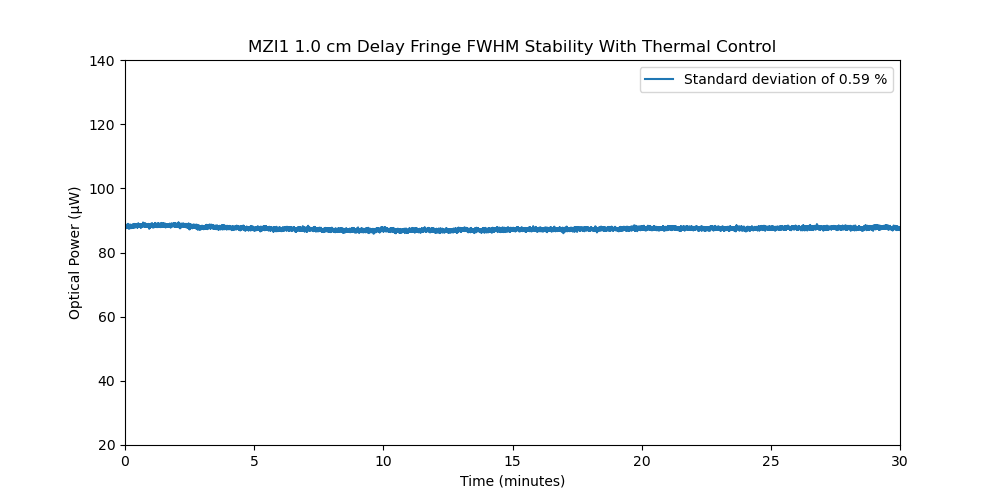}\end{tabular}
   \end{center}
   \caption[example] 
   { \label{fig:Flux With and Without Thermal Control} 
Flux at fringe inflection point over a 30-minute period without thermal stabilization (left) and with thermal stabilization (right), with a 1 cm delay path.}
   \end{figure}

For each configuration of delay path length, phase error was measured over 2-hour-long monitoring periods. Without any thermal stabilization in place, accurate computation of phase error was not possible due to thermal drift. Large changes in optical power over time were observed due to a slow global increase in chip temperature from heat dissipated by the MZI switches, as can be seen in the left plot of Fig. \ref{fig:Flux With and Without Thermal Control}. The increase in chip temperature affects both the phase and period of the fringes, as well as the switching behavior of each MZI switch. 

The same measurements were repeated with active thermal stabilization, which made computation of phase error metrics possible. The maximum phase error observed amongst all path length configurations was +/- 0.025 radians. No large fluctuations in optical power at the fringe inflection point were observed, as shown in the right plot of Fig. \ref{fig:Flux With and Without Thermal Control}. The phase error computed for each delay configuration is given in Table \ref{Phase Error Metrics}.

\begin{table}[h]
    \centering
 \caption{Phase error in radians and as a percent of fringe amplitude for each delay path length, measured over 2-hour-long periods.}
\label{Phase Error Metrics}
    \begin{tabular}{|c|c|c|c|c|} \hline 
         Path Length (cm)&  0&  0.5&  1& 1.5\\ \hline 
         Phase Error (rad)&  +/- 0.015&  +/- 0.025&  +/- 0.015& +/- 0.015\\ \hline 
         Phase Error (\%)&  +/- 1.46&  +/- 2.45&  +/- 1.51& +/- 1.52\\ \hline
    \end{tabular}

\end{table}

\section{CONCLUSIONS AND FUTURE WORK}

\subsection{Conclusions}
We demonstrate a system to precisely control the global temperature of a photonic chip with long delay lines, thus minimizing phase error originating from thermally induced changes to optical path lengths. The apparatus is constructed from inexpensive and readily available off-the-shelf components, exhibits high thermal stability of the order of +/- 0.028° C, and sub-minute settling time with minimal thermal overshoot/undershoot. This system successfully stabilized the interference fringes of a photonic Fourier transform spectrograph chip, reducing phase error to under +/- 0.025 radians for delay path lengths of up to 1.5 cm. 

\subsection{Potential Improvements}
While the stability achieved with this system is sufficient to carry out spectral reconstructions, further tuning and implementation of adaptive PID control algorithms would allow for more rapid stabilization of delay configurations and thus allow faster acquisition of spectra. In an adaptive PID loop, a more aggressive set of PID coefficients would be used when there is a large deviation between the actual chip temperature and the set temperature, while a less aggressive set of coefficients is used when the chip temperature is near the set point. Additionally, increasing the thermal mass of the heat sink coupled to the hot side of the TEC would improve the efficiency of heat transfer of the TEC, further improving the stability of the device.

\appendix    

\acknowledgments 
Support for P Gatkine was provided by NASA through the NASA Hubble Fellowship Grant
HST-HF2-51478.001-A awarded by the Space Telescope Science Institute, which is operated by the
Association of Universities for Research in Astronomy, incorporated, under NASA Contract NAS5-26555.
This work was supported by the Wilf Family Discovery Fund in Space and Planetary Science, funded by the Wilf Family Foundation, as well as the support from Keck Institute for
Space Studies at Caltech. Some of this research was carried out at Caltech and the Jet Propulsion Laboratory
and funded through the President’s and Director’s Research \& Development Fund program. This work was supported by NASA through the Center Innovation Fund.  

The authors would like to thank the staff at Lionix for the fabrication of the photonic chips presented herein.

\bibliography{report} 
\bibliographystyle{spiebib} 

\end{document}